\documentclass{iaus}
\usepackage{graphicx}
\usepackage{amssymb}
\def\eps@scaling{1.0}%
\newcommand\epsscale[1]{\gdef\eps@scaling{#1}}%
\newcommand\plotone[1]{%
 \centering
 \leavevmode
 \includegraphics[width={\eps@scaling\columnwidth}]{#1}%
}%
\newcommand\plottwo[2]{%
 \centering
 \leavevmode
 \columnwidth=.45\columnwidth
 \includegraphics[width={\eps@scaling\columnwidth}]{#1}%
 \hfil
 \includegraphics[width={\eps@scaling\columnwidth}]{#2}%
}%

\newcommand{\sun}{\odot}

\title[SEDs of High-Mass Star Formation] 
{2-D and 3-D Radiation Transfer Models of High-Mass Star Formation}

\author[Whitney et al.]   
{Barbara A. Whitney,$^1$ %
Thomas P. Robitaille,$^2$ R\'{e}my Indebetouw,$^3$ Kenneth Wood,$^2$ J. E. Bjorkman,$^4$
 Pia Denzmore$^5$}

\affiliation{$^1$Space Science Institute , 4750 Walnut St. Suite 205,
Boulder, CO 80301, USA \break email: bwhitney@spacescience.org\\[\affilskip]
$^2$School of Physics \& Astronomy, University of St Andrews, Scotland \break email: tr9@st-andrews.ac.uk, kw25@st-andrews.ac.uk\\[\affilskip]
$^3$Astronomy Department, University of Virginia, USA; 
email:remy@virginia.edu\\[\affilskip]
$^4$Ritter Observatory, University of Toledo, USA;
email: jon@physics.utoledo.edu\\[\affilskip]
$^5$Physics and Astronomy Department, Rice University, USA; email:piadenz@rice.edu}

\pubyear{2005}
\volume{227}  
\pagerange{000--000}
\date{00-00-2005 and in revised form 00-00-2005}
\jname{Proceedings Title IAU Symposium}
\editors{A.C. Editor, B.D. Editor \& C.E. Editor, eds.}
\begin{document}

\maketitle

\begin{abstract}
2-D and 3-D radiation transfer models of forming stars generally produce bluer
1-10 $\mu$m colors than 1-D models of the same evolutionary state and envelope
mass.  
Therefore, 1-D models of the shortwave
radiation will generally estimate a lower envelope mass and later evolutionary state
than multidimensional models.
1-D models are probably reasonable for very young sources, or longwave analysis ($\lambda >
100 \mu$m).
In our 3-D models of high-mass stars in clumpy molecular clouds, we find no correlation
between the depth of the 10 $\mu$m silicate feature and the longwave ($> 100 \mu$m) SED
(which sets the envelope mass), even when the average optical extinction of the envelope is
$> 100$ magnitudes.  This is in agreement with the observations of Faison et al.
(1998) of several UltraCompact HII  (UCHII) regions, suggesting that many of these sources
are more evolved than embedded protostars.

We have calculated a large grid of 2-D models and find substantial overlap between different
evolutionary states in the mid-IR color-color diagrams.
We have developed a model fitter to work in conjunction with the grid to analyze
large datasets.  This grid and fitter will be expanded and tested in 2005 and released
to the public in 2006.
 
\end{abstract}

\firstsection 
\section{Introduction}

An explosion of data is being collected at near- and mid-infrared wavelengths on star formation
regions throughout the Galaxy and beyond.  
Several key projects on the {\it Spitzer Space Telescope}  are imaging hundreds of
high-mass star formations regions at unprecedented resolution and sensitivity (Benjamin
et al. 2003, Marston et al. 2004, Megeath et al. 2004, Whitney et al. 2004a, Jones et al. 2005).
This combined with submillimeter and far-infrared imaging is providing complete SEDs of
several regions.

SEDs of massive star formation regions have been modeled extensively with 1-D 
radiative transfer codes  (Faison et al. 1998, van der Tak et al. 2000,
Beuther et al. 2002, Meuller et al. 2002, Williams, Fuller, \& Sridharan 2005).
The 1-D models do a reasonable job fitting the SEDs and radial intensity
profiles at wavelengths $> 100 \mu$m.
However, in all but the very youngest sources (Osorio, Lizano \& D'Alessio 1999), 1-D models that fit the longwave data almost invariably underestimate
the shortwave flux (1-20, sometimes 1-50 $\mu$m) (Mueller et al. 2002,
van der Tak et al. 2000, Williams et al. 2005) or require large inner radii or unusual
dust properties (Faison et al. 1998, Hatchell et al. 2000).

The shortwave radiation from Young Stellar Objects (YSOs) is much more susceptible to
geometric effects of the surrounding circumstellar material  than longwave radiation (Whitney et al. 2003a,b).   
Our approach is to estimate the most appropriate circumstellar geometries, based on theory
(Shu, Adams, \& Lizano 1987, Bonnell et al. 1998, Yorke \& Sonnhalter 2002, McKee \& Tan 2003)
and observations (Beuther et al. 2004, Beltr{\' a}n et al. 2004, Sandell \& Sievers 2004, van der Tak \& Menten 2005), and compute SEDs using 2-D and 3-D radiation transfer codes.
We show in this paper examples of calculated SEDs from an accretion scenario
(\S2) and from more complicated clumpy structures (\S3).
These 2-D and 3-D models have the common characteristic of allowing much more
shortwave radiation to escape than 1-D models, whether due
to bipolar cavities or clumpy geometries.   Unusual dust properties or large inner
dust radii are not required to simultaneously fit the shortwave and longwave SEDs.

With these multi-dimensional models comes more parameter space to explore in fitting
your favorite source.  
With this in mind we are in the process of computing a large grid of 2-D and 3-D radiation
transfer models.  We have developed a fitter that will use linear regression
to find the best model from the grid, and provide statistics on the uniqueness of
the fits (Robitaille et al. 2005, in preparation).   
This is discussed in \S4.



\section{The Accretion Scenario:  2-D Models}\label{sec:2d}

Our 2-D models have three components, in addition to a central illuminating source:  
a rotationally flattened infalling envelope, using
the solution for free-fall collapse of Ulrich (1976); a flared disk, assuming the hydrostatic
equilibrium solution (Shakura \& Sunyaev 1973; Lynden-Bell \& Pringle 1974); and partially evacuated bipolar cavities.
The models include accretion luminosity in the disk and an accretion hotspot on the star
with a temperature determined by the disk accretion rate (Calvet \& Gullbring 1998).
The models are described in several publications (Bjorkman \& Wood 2001,
Wood et al. 2002, Whitney et al. 2003a,b, 2004b).
The flared disk geometry is likely not strictly correct when embedded in an infalling envelope
but simulations have shown that the disks are vertically extended in infalling
envelopes (Bate 1998).
In addition,  observations of disks inside envelopes show similar properties to flared disks
(Wolf, Padgett, \& Stapelfeldt 2003).  The shape and density structure of the outflow cavities
is probably our biggest uncertainty.  We hope that simulations (e.g., Delamarter, Frank, \& Hartmann 2000) and observations will shed light on this.
The dust properties in our models can vary spatially.  We attempt to include appropriate grain models
for the disk, envelope, and cavity, as discussed 
in Whitney et al. (2003b).

To simulate objects in various evolutionary states we vary the envelope mass (infall rate);
disk mass and radius; and cavity size and density.   In more evolved sources,
the envelope mass is set to zero, leaving only a disk.

\begin{figure}
 \plotone{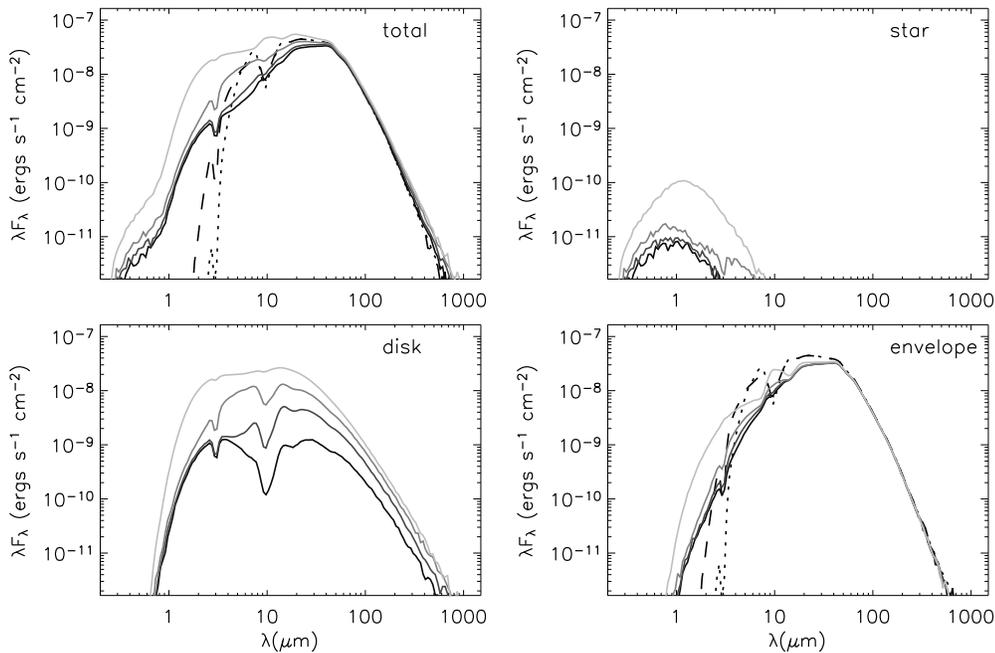}
  \caption{
 Model SEDs of a high mass B0 protostar.   Model parameters are described in the text. The top left panel shows the total SED.  The other panels show the contribution from photons whose last point of origin is from the star (top right),
 disk (bottom left), or envelope (bottom right).   Four inclinations are plotted as different shades of grey
 from edge-on (black) to pole-on (light grey).
 The dotted and dashed lines are 1-D models, described in the text.
      }\label{protosed1}
\end{figure}

\begin{figure}[t]
 \plotone{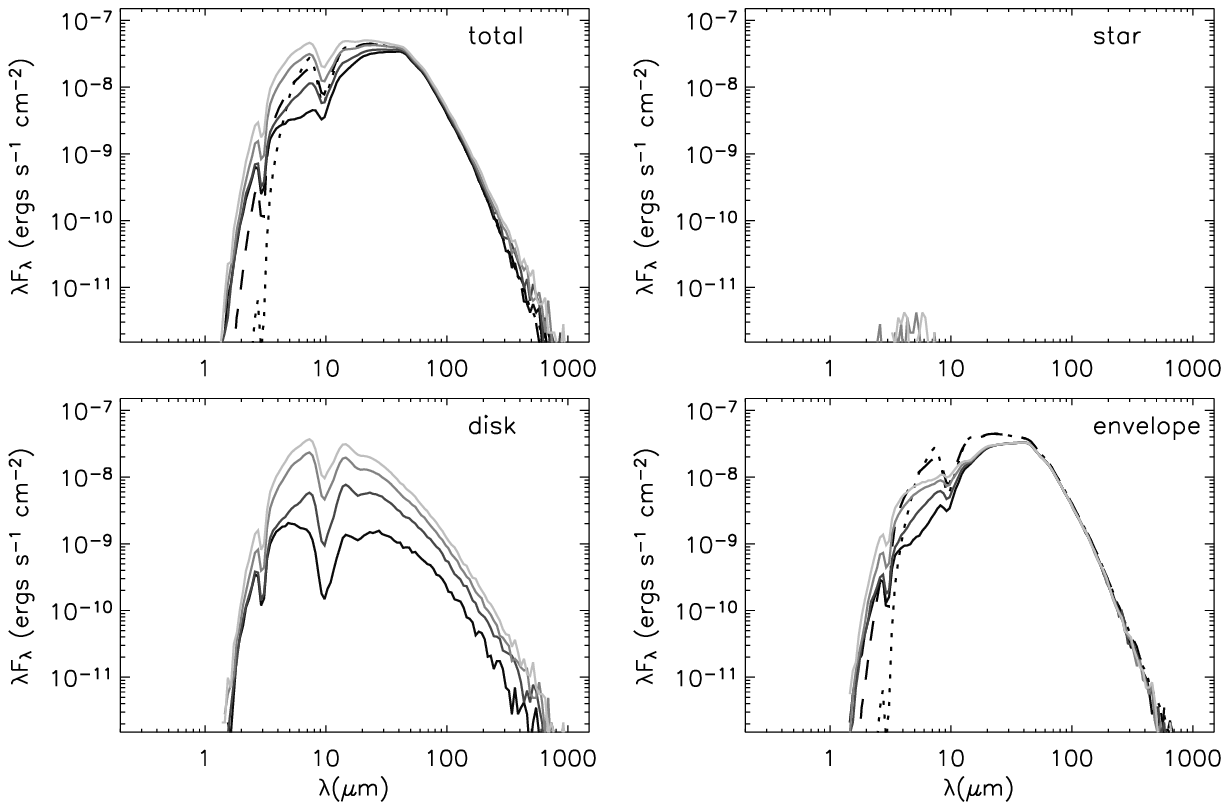}
  \caption{
  Same as Figure 1, but the 2-D model does not include a bipolar cavity.
      }\label{protosed2}
\end{figure}

A sample high-mass protostar model is shown in Figure \ref{protosed1}.   The central source is a B0 
main sequence star.  It is surrounded by a 500 AU radius disk of 1 $M_\sun$ mass, and
an envelope with an infall rate of $10^{-4}  M_\sun/$yr.   The centrifugal radius of the
envelope. $r_c$, is set to the disk radius of 500 AU.
The envelope has a bipolar cavity
carved out  with
an opening angle of 15 degrees at 30,000 AU; the cavity is filled with gas and dust with
a constant density of $6 \times 10^{-20}$ gm cm$^{-3}$ or $n_{H2} = 18,000$ cm$^{-3}$.
The spectra in Figure \ref{protosed1} are divided into four components based on the origin of the photons.
Ultimately all of the photons originated from the star, but when they are absorbed in the envelope
or disk and re-emited thermally, we say that this is their new point of origin.
Scattering does not count as a point of origin
 so each panel includes radiation that has scattered in the disk/envelope.
The top right panel shows photons that originated from the star and were not absorbed by
the circumstellar disk/envelope.  
The bottom left panel shows photons that originated from the disk, and the bottom right
shows those originated in the infalling envelope.  
In each panel the four solid lines show different inclinations in different
shades of grey, from edge-on (black) to pole-on (light grey).
The dotted and dashed lines show 1-D models for the same envelope mass.
The rotational infall model approaches a power-law solution of $\rho \sim r^{-\alpha}$,
with $\alpha=1.5$
at large radii.  
However, well within the centrifugal radius, $r_c$, $\alpha = 0.5$.  
The dotted line model has a single power-law for the density with $\alpha =1.5$.
The dashed line model has $\alpha = 0.5$ for $r < r_c$ and $\alpha = 1.5$ for $r>r_c$.
This allows more shortwave radiation to escape but not nearly as much as the 2-D model.
To get more shortwave radiation in the 1-D model would require lowering the envelope mass.
This would still produce too steep a slope in the rising part of the SED, and would no longer fit the longwave SED.  In short, the 1-D
models can not simultaneously fit the entire SED of a 2-D geometry.

A few things to note about these spectra are 1), The disk is a significant contributor to
the SED shape at $\lambda < 100 \mu$m in the 2-D model.  This is partly due to the fact
that it is dense and warm in the inner regions, and that the bipolar cavity allows the
radiation to escape.  And 2), the 1-D model is
not that different from the envelope component of the 2-D model.
Figure \ref{protosed2} shows the same 2-D model without a bipolar cavity.   Here the 1-D models,
especially the 2-component power-law model,
are in better agreement with the 2-D model.
The disk SED is narrower than in the cavity model (Figure \ref{protosed1}) because more of the shortwave photons are reprocessed by the envelope.
Since the cavity shape, size and density is uncertain, this adds uncertainty to the 2-D models.
However, outflows and therefore cavities appear to be ubiquitous in even the youngest protostars,
and cavities may be a necessary component to allow high mass stars to form (Krumholz, McKee, \& Klein 2005), so it is reasonable to include them.

Examples SEDs of disk models can be found in several of our previous publications 
(Wood et al. 2002, Whitney et al. 2003b, 2004b) as well as those of several other 
groups (Chiang \& Goldreich 1997, D'Alessio et al. 1998, 1999, 2001; 
Dullemond, Dominik, \& Natta 2001, Dullemond \& Dominik 2004), and are not shown here.

%


\section{3-D Clumpy Models for UCHII Regions}

Our clumpy models are motivated by previous SED models of UCHII regions
which have a common feature that to fit the longwave spectra,
the 1-D models tend to underestimate the shortwave flux and to give too deep a 10 $\mu$m
silicate feature (Faison et al. 1998).
2-D and 3-D models that allow a path of escape for the shortwave photons can solve this problem.

UCHII regions are thought to be O or early B stars in a later stage of evolution than the
Hyper-Compact H II regions or the hot molecular cores (Churchwell 2002).
These high-mass stars are so luminous that they heat up large volumes 
of the interstellar medium (ISM).  
For example, an O5 star will heat dust up to 30 K out to a radius of 0.75 pc from the
star.  
Thus the clumpy nature of the ISM may be important in addition to the nearby circumstellar
geometry.
Because the UCHII regions may be in a later evolutionary state than other high mass protostars, 
we chose to consider a simple model of an O
star embedded in a clumpy molecular cloud (Indebetouw et al. 2005a).
To approximate the density structure of the molecular cloud, we assume a
hierarchically clumped structure with a fractal dimension of 2.6, as estimated
from observations (Elmegreen \& Falgarone 1996).
The prescription of Elmegreen (1997) produces the fractal density structure.
This structure varies depending on the choice of an initial random number seed.
We place a star in the center of this clumpy gas and dust, and include a smooth component along with the fractal density.  
For the models presented here, the fraction of the mass in the smooth component
is 10\%.  

\begin{figure}
{\centerline{
{\includegraphics[width=4.5in]{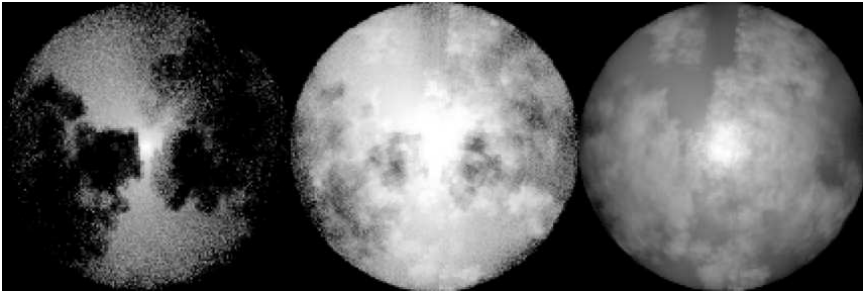}}}}

  \caption{
Images from a single viewing angle of a clumpy model at 1-3 $\mu$m (left), 3-8 $\mu$m (middle) and
24-160 $\mu$m (right).
       }
\label{image}
\end{figure}

\begin{figure}
\epsscale{0.9}
 \plotone{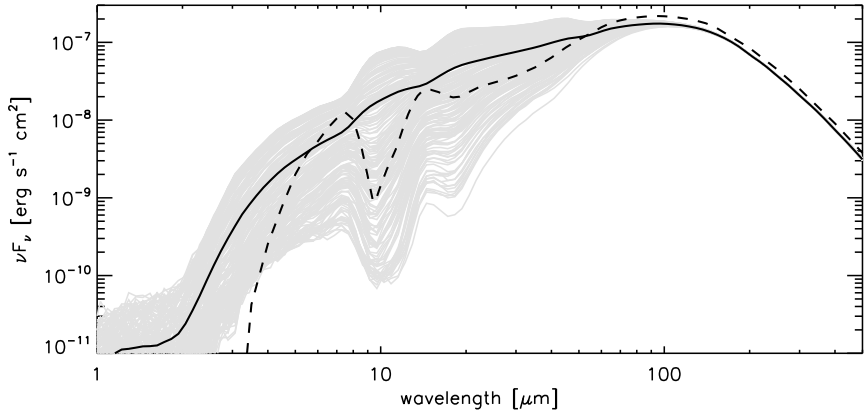}
  \caption{
The grey lines show 200 sightlines from a single 3-D clumpy model. 
The solid line is the average of the 200 sightlines.
 The dashed line shows
a 1-D model for the same envelope mass.  
       }\label{clump1}
\end{figure}

Figures \ref{image} and \ref{clump1} show the results from a sample model of  an O5 star (T=41000 K) illuminating a clumpy cloud.
Figure \ref{image} shows images of the cloud and Figure \ref{clump1} shows the SEDs.
The inner radius is the dust destruction radius; the outer is 2.5 pc.
The optical extinction through the envelope ranges from A$_V =$ 13 to 401, depending on viewing angle, with an average of 131.  
The grey lines in Figure \ref{clump1} show 200 sightlines from this model.  The
solid line shows the average; and the dashed line is the result from a smooth (1-D) model
with the same total mass and average radial density profile (in this case, constant density).  
Notice the large variation in the 10 $\mu$m silicate feature with viewing angle,
and the convergence of all the viewing angles for $\lambda > 100 \mu$m.
While the longwave SED points to a large envelope mass (50,000 $M_\sun$ within
a 2.5 pc radius), the silicate feature can
be in emission for viewing angles which happen to have a low line-of-sight extinction.
This is similar to observations of UCHII regions (Faison et al. 1998).

In our exploration of parameter space to fit the data of Faison et al. (1998), we
varied the smooth-to-clumpy mass ratio from 0-100\% (pure clumpy to smooth),
the spherically-averaged radial density
power law exponent from -2.5 to +1, 
and the fractal dimension from 2.2 to 3.
Our best fit models have a smooth-to-clumpy ratio of 10\% and a radial density
profile exponent of 0 (Indebetouw et al. 2005a).

\begin{figure}
\epsscale{0.9}
 \plotone{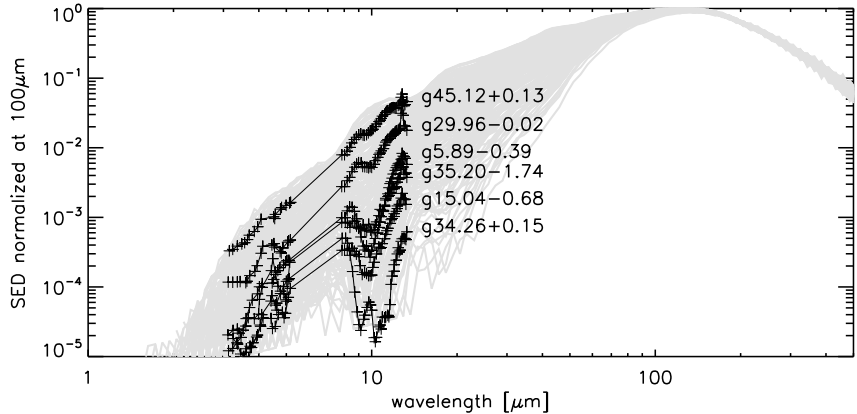}
  \caption{
 All of the Faison et al. (1998) spectra, normalized to their 100 $\mu$m values are plotted along with 
 all the viewing angles from a single clumpy model.  
        }\label{alldata}
\end{figure}

Figure \ref{alldata} shows the set of IR spectra of UCHII regions from Faison et al. (1998)
normalized to the 100 $\mu$m value.  These are overlayed on the 200 sightlines
from a single clumpy model. 
The silicate feature in the data has very similar behavior to the 3-D models.
In 1-D models the depth of the silicate feature is correlated with envelope mass.  In
the 3-D models and these data, there is no such correlation.



The fact that these UCHII region spectra are well fit by a massive star in a molecular
cloud, with a flat density profile, suggests that they are indeed more evolved
than a protostar surrounded by an infalling envelope.
We plan to model individual sources in the future, taking into account source illumination
and geometries suggested by observations.

\begin{figure}[ht!]
\epsscale{1.0}
 \plotone{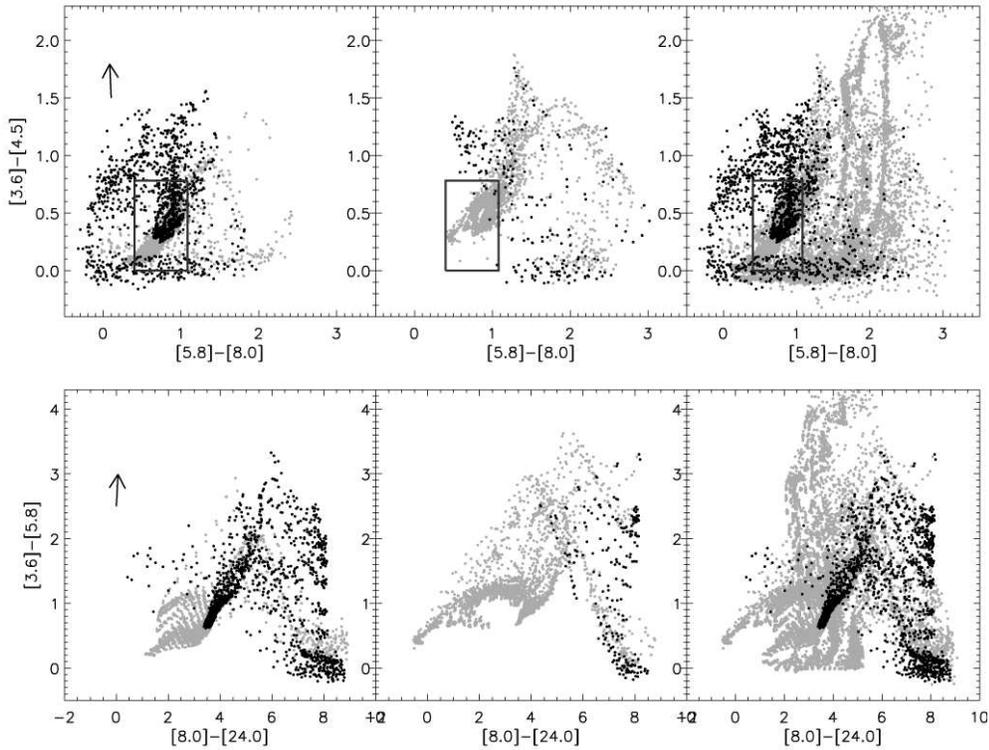}
  \caption{
Color-color plots of the model grid.  The grey points are disk-only sources;
the black point are younger disk+envelope sources (protostars).  The arrows show reddening vectors (A$_V=20$) based
on a standard interstellar extinction law.  The grey box is the Allen et al. (2004) disk domain.
Top: objects with $M_\star < 2 M_\sun$ and inner radii at the dust destruction
radius.  Middle: objects with $M_\star > 2 M_\sun$ and inner radii at the dust destruction
radius.  Bottom:  all stellar masses and including models with inner holes of various sizes.
       }\label{ccall}
\end{figure}

\section{A Model Grid and Fitter}

As described in the introduction, the {\it Spitzer Space Telescope} and other observatories
are imaging a large number of star formation regions.  The GLIMPSE project alone
(Benjamin et al. 2003, Indebetouw 2005b) has imaged 220 square degrees of the inner galactic plane
and produced catalogs of over 40 million sources.  
To help analyze these and other data sets, we are producing a large grid of 2-D and 3-D YSO
models (Robitaille et al., in preparation).
Our current grid contains 1600 models.  Since each model produces SEDs for 10 viewing
angles, the total number of spectra is 16,000.

Figure \ref{ccall} shows mid-IR color-color plots of the model grid.
The top panels show colors in the {\it Spitzer} IRAC bands (3-8 $\mu$m; Fazio et al. 2004) ,
and the bottom panels include the MIPS 24  $\mu$m band (Rieke et al. 2004).
The left panels show low mass stars ($< 2 M_\sun$).  The grey points are sources with
disks with inner radii set to the dust destruction radius; the black points are younger sources
with envelopes in addition to disks.   We will refer to these younger sources with envelopes
as ``protostars'', and the more evolved disk sources as ``disk'' sources.  
The protostars are plotted on top of the disk sources so the black points cover
up some of the
grey points, especially in the boxed region.  This boxed region in the top panels
was denoted by Allen et al. (2004)
as the disk domain, based on their SED models.  
Since the model grid has a continuum of evolutionary states, we choose a dividing line
betweeen protostars and disks at an infall rate where the envelope is relatively low-mass and tenuous.
For these plots, we choose the dividing line to be 
an infall rate of $\dot M = 10^{-7} M_\sun/$yr $\times M_\star/M_\sun$.   
The protostars tend to be bluer in [5.8]-[8.0] than those presented by Allen et al.
This is likely due to our inclusion of bipolar cavities which make the sources more blue
at these wavelengths, as Figures \ref{protosed1} and \ref{protosed2} show.   
In fact, the protostars tend to be bluer than the disks at [5.8]-[8.0].  
The protostars are redder 
than most of the disks at [8.0]-[24] (bottom left panel) as expected.

The middle panels of Figure \ref{ccall} shows color-color plots for sources with stellar mass greater that 2 $M_\sun$
and inner radii at the dust destruction radius.
These show a larger spread in colors than the low-mass sources, with a redder tail at
[5.8]-[8.0].
This is likely due both to   
higher stellar temperatures and the larger spatial scale in the circumstellar temperature
profile in the high-mass sources (Whitney et al. 2004b).
These panels also show that we didn't include enough high mass protostars in our grid
which will be rectified in the near future.
As in the low-mass sources, the protostars are redder at [8.0]-[24] than most of the disks.  

The left and middle panels of Figure \ref{ccall}  show some separation in evolutionary stage with
color and perhaps mass, which is encouraging.
The right panels show an interesting feature when disks with inner holes are included.
These are redder in the {\it Spitzer} IRAC bands than even most of the protostars,
due to a lack of hot dust.   
Even very low mass disks or remnant dust surrounding luminous sources can be quite
bright and have very red colors.   These should not be mistakenly interpreted as embedded
protostars based on their red IRAC colors.
The bottom right panel shows that these disks are reasonably well-separated from 
protostars when the [24] $\mu$m data point is included, though there is a fair amount of overlap
(the protostars are overplotted on the disk sources and hide some of them).


\begin{figure}[t!]
\epsscale{0.8}
 \plotone{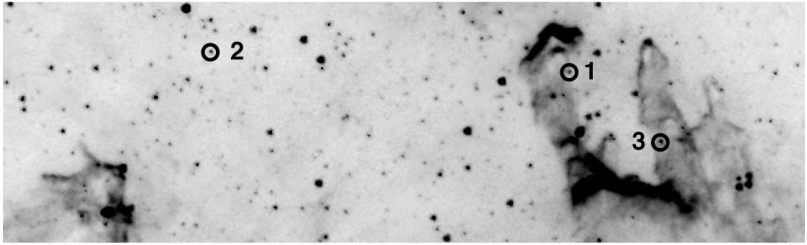}
  \caption{
GLIMPSE 8 $\mu$m image of pillars in the M16 star formation region.  The three sources modeled in Figure \ref{m16seds} are labeled by numbers 1-3.
       }\label{m16}
\end{figure}

\begin{figure}
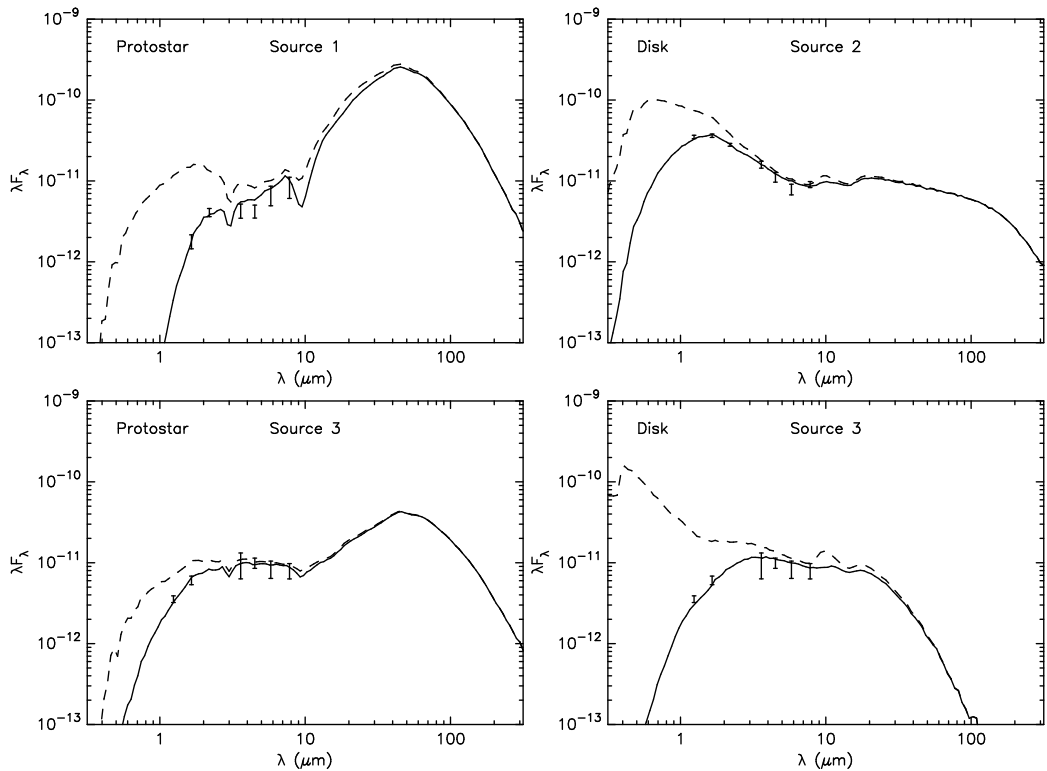

{\rotatebox{270}{\includegraphics[width=2.in]{whitney_b_fig8a.ps}\includegraphics[width=2.in]
{whitney_b_fig8d.ps}}}
{\rotatebox{270}{\includegraphics[width=2.in]{whitney_b_fig8b.ps}\includegraphics[width=2.in]
{whitney_b_fig8c.ps}}}
  \caption{
Data and models for the three sources labeled in Figure \ref{m16}.  The 2MASS and IRAC data are plotted with error bars.  The best model fit including extinction is shown as a solid line; the same model without extinction is shown as a dashed line.  The model parameters are described
in the text.
 }
\label{m16seds}
\end{figure}

We have developed a fitting program that uses linear regression to find the
model spectrum that best agrees with an observed SED.
The model SEDs are convolved with the broadband filter
functions of the dataset (any can be included) before the fitter program is run.
The fitting program produces statistics on quality of fits and uniqueness.
When the wavelength range being fit is narrow (for example, 2MASS\footnote{This publication makes use of data products
from the Two Micron All Sky Survey, which is a joint project of the
University of Massachusetts and IPAC/Caltech, funded by NASA and the
NSF}/IRAC data
at 1-8 $\mu$m),
a single observed SED can have 100 good fits, so we produce histograms of
model parameters that fit a given source.
The fitter simultaneously fits a foreground extinction along with the best SED fit.
Thus the program can also be used to map extinction in a region, as shown in
Indebetouw (2005b).
Knowing the distance to an observed cluster improves the model fits by better constraining
the luminosity of the sources and narrowing the range of good fits.  This also improves
the stellar mass estimates.

Figures \ref{m16} and \ref{m16seds} show some results of our model fitter.
The GLIMPSE IRAC and 2MASS SEDs of three sources identified in Figure \ref{m16} are 
plotted in Figure
\ref{m16seds} along with the best model fits.
Source \#1 is well fit by a protostar with $M_\star = 1.9 M_\sun$, $T_\star$ = 4200 K,
an envelope infall rate of $10^{-4} M_\sun$/yr, and foreground extinction A$_V = 13$.  
Source \#2 is well fit by a star with $M_\star = 3 M_\sun$, $T_\star$ = 5000 K, a disk with mass of $0.04 M_\sun$ and outer radius of 300 AU, and A$_V=3$.   
The third source (bottom panels) did not produce a unique fit.
The bottom left panel shows a protostar model for this source with 
$M_\star = 1.2 M_\sun$, $T_\star$ = 4050 K,  envelope infall rate 
$4\times 10^{-6} M_\sun$/yr, and A$_V=3$.   
The bottom right panel shows a disk model for this source with 
$M_\star = 2.5 M_\sun$, $T_\star$ = 10,500 K, disk mass of $6 \times 10^{-5} M_\sun$,
and A$_V=8$.
Often the ambiguous fits trade off stellar temperature and foreground extinction with evolutionary state (i.e., a cooler $T_\star$ protostar with lower A$_V$ has a similar 1-10 $\mu$m spectrum as a hotter $T_\star$ disk source with a larger A$_V$).
Clearly, longer-wave observations would distinguish between these two models.  Also, note the large range of A$_V$ between the two models.  

\section{Conclusions}\label{sec:concl}

The process of forming a star is at a minimum a 2-dimensional process.  In the accretion
scenario, angular momentum
conservation produces a rotationally flattened envelope and a disk.  Some combination
of angular momentum and magnetic fields produces bipolar jets and outflows very early on 
(Shepherd, 2005 and references therein).
And turbulence and various dynamical effects produces clumps and other asymmetric structures.
We've shown that 2-D and 3-D SED models can produce bluer near- and mid-IR colors than 1-D models;
and the depth of the 10 $\mu$m silicate feature is uncorrelated with the envelope mass
in some clumpy geometries.
We believe that multi-dimensional models are required to interpret the shortwave observations in all but
the youngest, most embedded sources.
To aid in the interpretation of the data, we are producing a large grid of models.
This meeting provided a wealth of useful information that will be used to improve
our grid of models.  
After our model fitter has been tested, we will make it publicly available by web access.
Users will be able to submit datasets as batch jobs.
We hope to have this set up by early 2006.

\begin{acknowledgments}
We are grateful to Ed Churchwell and John Mathis for many useful discussions on
high mass star formation and clumps.
This work was supported in part by a NASA LTSA grant NAG5-8933 (BAW), 
a {\it Spitzer} Fellowship (RI), and a UK PPARC Advanced Fellowship (KW).
\end{acknowledgments}



\end{document}